# A. Bettini

G. Galilei Physics Department. Padua University
INFN. Padua
Canfranc Underground Laboratory
Bettini@pd.infn.it

# THE WORLD UNDERGROUND SCIENTIFIC FACILITIES
## A COMPENDIUM

**Abstract**. Underground laboratories provide the low radioactive background environment necessary to explore the highest energy scales that cannot be reached with accelerators, by searching for extremely rare phenomena.

I have requested to the Directors of the Laboratories a standard set of questions on the principal characteristics of their laboratory and collected them in this compendium.

I included the ideas and plans for short-range developments. However, next-generation structures, such as those for megaton-size detectors, are not discussed.

A short version of this work will be published in the Proccedings of TAUP 2007.

# Preface

This document is meant to provide a summary of the principal physical characteristics of the Underground Physics Laboratories worldwide. I prepared it on the basis of the information I collected, in a standard format, from the Directors in July-August 2007. It is a pleasure to thank Eugenio Coccia, Fedor Danevich, Fred Duncan, Hiro Ejiri, Vladimir Gavrin, Sun-Kee Kim, Kevin Lesko, Naba Mondal, Tony Noble, Juha Peltoniemi, Earl Peterson, Fabrice Piquemal, Neil Spooner, and Yoichiro Suzuki for their kind help. Further information can be found in the web sites quoted under each laboratory.

I have included all the existing infrastructures, known to me, deeper than about 500 m and hosting astroparticle physics experiments. I included also the two existing projects of new laboratories, which have reached a fairly advanced stage in the approval process, INO in India and DUSEL in the US. Other ideas exist, but being they at a preliminary stage, have not been included.

There are important differences amongst the infrastructures, not only in their physical characteristics, like depth, dimensions, type of access, environment, etc. They may be handled as an international laboratory or observatory with an international users community, services and an International Scientific Committee helping the Director in the definition of the scientific programme or, at the other extreme, to be the underground facility of a specific group or collaboration of scientists, which defines its research programme.

The muon flux decreases with the thickness of the rock overburden roughly, but not exactly, exponentially. Under a flat surface the flux is $10^{-3}$ $m^{-2}s^{-1}$ at a depth of 2.3 km w.e. (1 km water equivalent is about 300 m of rock), $10^{-4}$ $m^{-2}s^{-1}$ at 3.7 km w.e. and $10^{-5}$ $m^{-2}s^{-1}$ at 5.3 km w.e. Under a mountain, like at Kamioka or Gran Sasso, the angular dependence of the muon flux is complicated, due to the shape of the surface. It must be measured to provide the input to the background simulations needed by the most delicate experiments. The flux is time dependent with seasonal variations of several percent.

Neutrons originate mainly from ($\alpha,n$) and fission processes (U and Th) in the rocks. The energies range from thermal to several MeV and consequently are not difficult to shield. The neutron flux is substantially independent of depth (if >100 m or so). It depends on the local geology. However, in practice the *n* flux is around a few $10^{-2}$ $m^{-2}s^{-1}$ in all laboratories.

Muons interactions in the rocks produce dangerous neutrons at a depth dependent rate. Fluxes are typically 3-4 orders of magnitude smaller than the main *n* flux. The energies are large, up to several GeV, demanding thicker shields.

Even more dangerous are the neutrons produced in the shields, in the detector and in the materials around it. In case of fast reactions, namely if the neutron immediately follows the muon, the background can be reduced by anticoincidence. Neutrons from metastable nuclides can be eliminated working at larger depth. Their effect depends on the details of the experiment and is generally more severe for high-Z materials. Clearly, the sensitivity to each background depends on the experiment. Consider for example the neutrons always produced by the muons in a liquid scintillator detector. Once moderated in some 250 $\mu$s they almost always end up by being captured by a proton with the emission of a 2.2 MeV gamma. Experiments searching for neutrino-less double beta decay with $Q_{\beta\beta}$ at or below this energy need to take particular care of this problem.

Radon ($^{222}$Rn) is a radioactive, volatile gas always present in the atmosphere, being continuously produced by the decay of $^{226}$Ra present in the rocks. An important source of Rn is ground water. Rn activity, which is typically 10 – 20 Bq/m$^3$ in open air, is larger by two orders of magnitude or more in closed underground cavities. It is reduced by ventilation. The equilibrium activity depends on the emanation rate and on the ventilation speed. The input air duct, which may be a few km long, should be made of SS or similar materials to avoid Rn collection input.

All experiments need to be shielded. Indeed, the shield thickness determines the physical size of the most sensitive experiments. For example, the frontier double beta decay experiments presently under construction, CUORE and GERDA, have diameters of about 15 m. Next generation may need halls 20-30 m in diameter and height. Notice that the maximum safe diameter decreases while the costs increase with depth.

Laboratories differ in many other important aspects: access horizontal or vertical, interference with nearby activities (mine work, road traffic, etc.), quality of the support infrastructures (laboratories, office space, assembly halls, etc.) and personnel on the surface, degree of internationality of the user community and programme of the advisory committee, policy of space and time allocation. Scientific sectors different form astroparticle physics as biology, geology and engineering can profit of the very special underground environment provided by the laboratories and their facilities. I shall not discuss these important issues here.

Design studies for a next-generation underground megaton size detector are under way in all the region of the world. Several underground laboratories are active in this effort. Since the scale of such projects is much larger than that of the existing laboratories, the issue requires a separate discussion and is not included in the present review.

The Particle and Nuclear Astrophysics and Gravitation International Committee (PaNAGIC) (http://www.iupap.org/wg/panagic/index.html) is a Working Group, WG4, of IUPAP created in 1998 to support international exchange of ideas and help in the convergence of the international scientific community in the large scale activity in the emerging field of particle and nuclear astrophysics, gravitation and cosmology. In particular, Underground Laboratories are included in its charge.

## EUROPE

The largest number of underground infrastructures is in Europe, being present both in its Western and Eastern parts. The largest sites are under a mountain, others are in a mine. They are here listed in alphabetical order.

ApPEC: Astroparticle Physics European Coordination, http://appec.in2p3.fr/. In 2001 several European scientific Agencies signed an agreement aiming essentially at

- promoting and facilitating co-operation within the growing European astroparticle physics community
- developing long term strategies and offering advice to national funding agencies or other institutions as appropriate
- expressing the views of the astroparticle physics community in international forums
- establishing a system of peer-review assessment applicable to projects where ApPEC members are involved.

Co-ordination actions have been set in place for the European Union laboratories in the 6$^{th}$ Framework Programme within the ILIAS project. http://www-ilias.cea.fr/

The proposal for its follow-on in the 7$^{th}$ Framework Programme is in preparation.

ASPERA, http://www.aspera-eu.org/index.php?option=com_frontpage&Itemid=44 is an action correlated with ApPEC, funded in the ERA-NET frame of the 6$^{th}$ Framework Programme

## BNO. Baksan Neutrino Observatory (Russia)

The Laboratory is operated by the INR of the Russian Academy of Sciences. It is managed as an observatory, with very long duration experiments. It is the oldest facility in the world built specifically for scientific research. Thanks to very wise and persistent policy of Moissey Markov, who was for a long time the Head of Nuclear Physics Division of the Academy of Sciences of the USSR, a special Decree of the Soviet Government was issued in 1966, and construction of the Baksan Neutrino Observatory started. The scientific activity started under the leadership of Alexander Chudakov and George Zatsepin.

**Director**: Valery Kuzminov

**Underground area available for experiments (including the already occupied one)**

a. The hall of the Baksan Underground Scintillation Telescope (BUST); volume $(24 \times 24 \times 16)$ m$^3$

b. The Laboratory of the Gallium Germanium Neutrino Telescope (SAGE); volume $(60 \times 10 \times 12)$ m$^3$

c. The construction of a larger and deeper hall for solar neutrino research, about 40 000 m$^3$ in volume, was started in 1990 and stopped in 1992, when the Soviet Union collapsed. Only a small fraction has been excavated. Further fate of this not finished construction is under discussion now.

d. Low Background Chambers with volume from 100 to 300 m$^3$ are used for R&D of dark

matter and double beta decay search as well as for gravitational wave search and for some geophysics measurements.

**Structure of the surface campus, general services provided to the users, number of employees of the lab**

A new village, called 'Neutrino', was built as a part of the original project in a previously empty space with personnel providing all necessary services (heating station, water supply system, first medical help, transportation, safety, etc.). The staff directly related to science is 50-60.

**Number of scientific users**: 30-35

**Web address**: http://www.inr.ac.ru/INR/Baksan.html

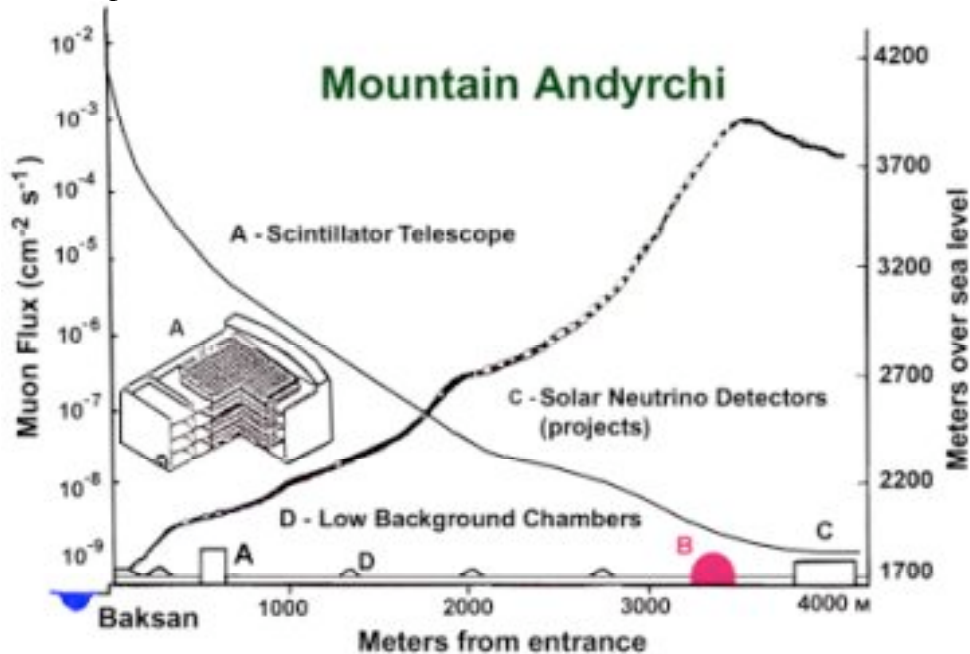

Fig. 1. Cross section of mount Andyrchi with the locations of main underground and surface facilities of the INR-RAS

**Type of access. Interference with nearby activities (freeway traffic, mine works, etc.)**

The underground laboratories are situated along two parallel horizontal tunnels each 4 km long excavated into the side of Mt. Andyrchi. Rail transportation underground (electric locomotives, passenger wagons, and special platforms) is used for personnel, materials, and equipment. So, there are no problems with nearby activities.

**Rock coverage (metres of rock), muon flux, neutron flux, radon activity (average with ventilation)**

BUST is located at an effective depth of 0.85 km w.e.

The SAGE laboratory is 2100 m deep (4.7 km w.e.); its volume is 7000 m$^3$, its area on the floor 720 m$^2$. Thickness of low background concrete =60 cm. Muon flux is $3.03\pm0.19 \times 10^{-5}$ m$^{-2}$s$^{-1}$ (average energy 381 GeV); Neutron flux (>1 MeV)=$1.4\pm10^{-3}$ m$^{-2}$ s$^{-1}$; (>3 MeV) =$6.28\pm10^{-4}$ m$^{-2}$ s$^{-1}$. Rn activity in air: 40 Bq/m$^3$.

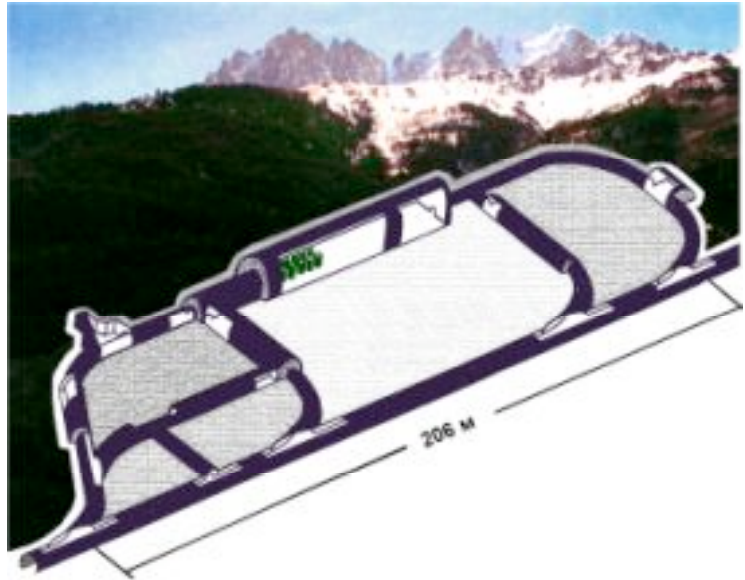

Fig.2

**Ventilation power (time to change one lab volume)**

Fresh air input ventilation ~ 60 000 $m^3$/h or 7 lab volumes/h; System of air purification: 2-stepped filters.

**Installed electrical power and available heat exhaust per unit time**

1450 kW. Heat is exhausted through ventilation air.

# BUL. Boulby Palmer Laboratory (UK)

The site was developed starting in 1998 by Neil Spooner and collaborators initially from Rutherford & Appleton Laboratory.

**Director**: Neil Spooner; n.spooner@sheffield.ac.uk

**Web address**: http://astro.ic.ac.uk/Research/ZEPLIN-III/boulby.htm

**Underground area available for experiments (including the already occupied one)**: About 1500 m$^2$

**Status of the occupancy of the above area and dates foreseen for the conclusions of the experiment**

ZEPLIN II complete by 2007/8; ZEPLIN III complete by 2010; DRIFT II ongoing R&D. Low radioactivity measurements. Geophysics.

**Structure of the surface campus, general services provided to the users, number of employees of the lab**

Surface building (200 m$^2$): computing, electronics, chemical labs, offices, conference room, changing rooms, mess rooms, mechanical workshop, storage and construction rooms. A workshop in a separate building. 2 people employed by the lab plus 1 from Sheffield University

**Number of scientific users**: about 30

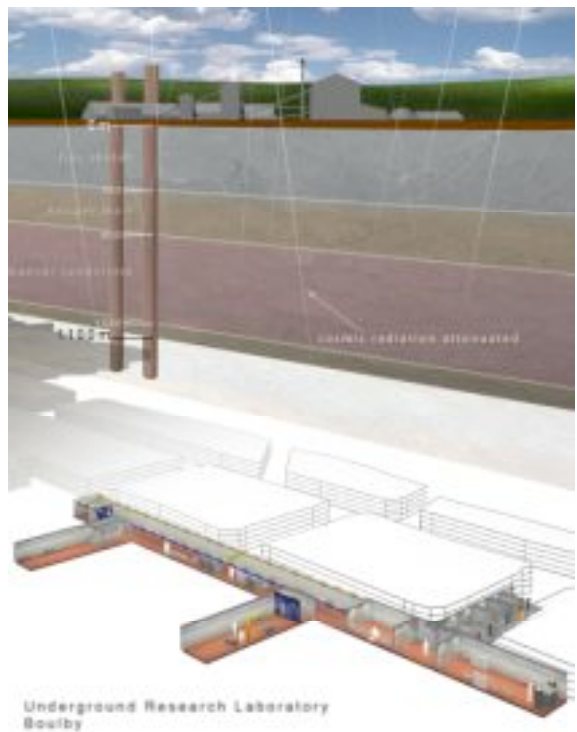

Fig. 1. The Underground Research Laboratory at Boulby

**Type of access. Interference with nearby activities**

Vertical through shaft of a working potash mine: transit time for people from surface lab to underground lab about 20'-25': 24 hr access with restrictions according to mine operations

**Rock coverage (metres of rock), muon flux, neutron flux, radon activity (average with ventilation)**

Rock overburden: 1000 m under flat surface (2.8 km w.e.)

Neutron flux: above 0.5 MeV: $1.7 \times 10^{-2}$ s$^{-1}$m$^{-2}$. Muon flux = $4.5 \times 10^{-4}$ m$^{-2}$ s$^{-1}$.

**Future**. Studies for the development of possible new areas in the salt layer and possibly in lower hard-rock areas.

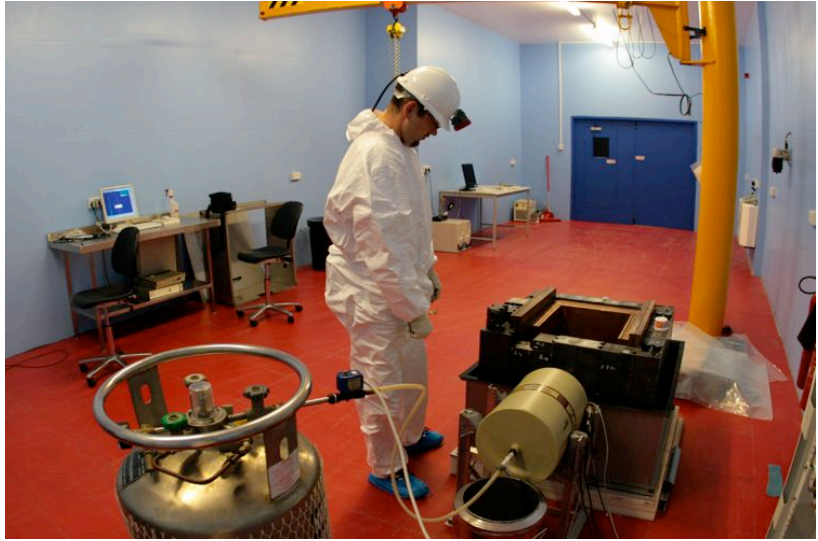

Fig. 2. One of the underground laboratories t Boulby

# CUPP. Centre for Underground Physics at Pyhäsalmi (Finland)

**Director**: Juha Peltoniemi juha.peltoniemi@oulu.fi

**Web address**: http://cupp.oulu.fi/

**Underground area available for experiments (including the already occupied one)**:

>1000 m$^2$ spaces no longer used by the mine

**Status of the occupancy of the above area and dates foreseen for the conclusions of the experiments**

EMMA experiment occupying 500 m$^2$ for a decade.

**Structure of the surface campus, general services provided to the users, number of employees of the lab**

Small lab and office in a surface building, other test laboratory and guest house.

About 3 FTE on site + 3 in Oulu university.

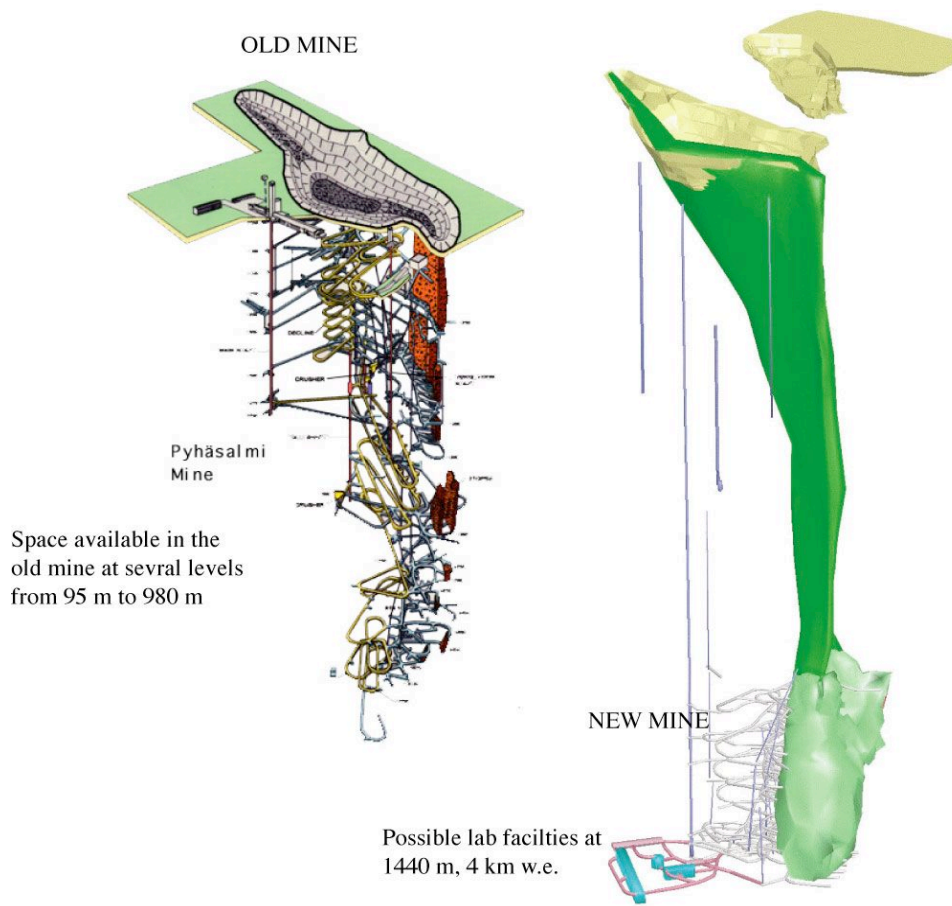

Fig. 1. The old and the new mine, showing the site for possible deep facilities

**Number of scientific users**: The members of the EMMA collaboration, about 20

**Type of access. Interference with nearby activities**

Inclined tunnel (nights closed); hoist to 1440 m level, pre-defined schedules. Must respect the operation of mine.

**Rock coverage (metres of rock), muon flux, neutron flux, radon activity (average with**

**ventilation)**

Space available at different depth, down to 1400 m of rock. Background depends on the site. Background measurements being performed.

**Installed electrical power and available heat exhaust per unit time**

Several kW available

**Ideas for the future**

Small to medium-scale experiments can be hosted in existing cavities. There are discussions on several possible experiments, which are at different phases of R&D. However, at the moment there no sufficiently supported proposal has been submitted.

# LNGS. Laboratori Nazionali del Gran Sasso. L'Aquila (Italy)

LNGS is a national laboratory of the Italian Istituto Nazionale di Fisica Nucleare (INFN). It is the largest in the world, serving the largest and most international scientific community.

In 1979 the President of the INFN A. Zichichi proposed to the Parliament to build a large underground laboratory close to the Gran Sasso freeway tunnel then under construction (an opportunity that reduced substantially the cost). In 1982 the Parliament approved the construction, which was completed by 1987. LNGS is operated as an international laboratory. In the process of approval of the proposals, an international Scientific Committee, appointed by INFN, advises the Director. Underground space and ancillary resources are allocated to experiments for a definite amount of time, in order to guarantee turnover.

Visits and tours in the laboratory are organised on a regular basis

**Director**: E. Coccia; coccia@lngs.infn.it

**Web address**: http://www.lngs.infn.it/

**Underground area available for experiments (including the already occupied one)**:

Three main halls, about $100 \times 20 \times 18$ (h) m$^3$ plus ancillary tunnels providing space for services and small-scale experiments. Two 90 m long tunnels for optical interferometer for geology studies. Total area 17 300 m$^2$, total volume 180 000 m$^3$.

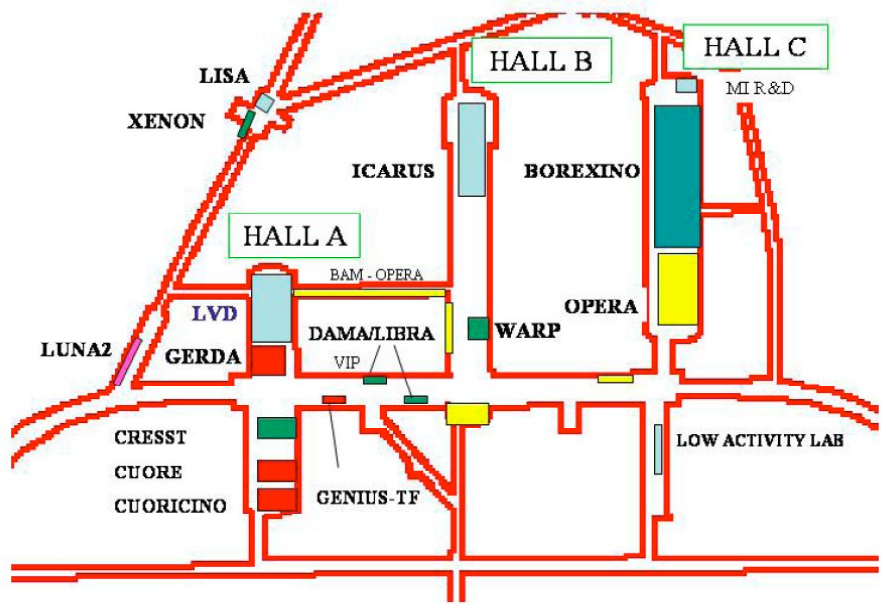

Fig. 1. Present occupancy

**Status of the occupancy of the above area and dates foreseen for the conclusions of the experiments**

**Dark matter**: LIBRA, CRESST2, XENON10, WARP; **Double Beta Decay**: COBRA, CUORICINO, GERDA; **Solar neutrinos (geo-neutrinos)**: BOREXINO; **Supernova** **neutrinos**: LVD; **Nuclear astrophysics**: LUNA2; **CNGS**: OPERA, ICARUS; **Equivalence principle:** VIP; **other**: R&D for LISA. **Low radioactivity measurements**

Other experiments on geology, biology, environmental issues, including two 90 m long Michelson interferometers for geological studies (not shown in Fig.1)

Almost all the experiments are second generation ones and are approved for several years of data taking.

**Structure of the surface campus, general services provided to the users, number of employees of the lab**

Services hosted on the surface campus include offices, mechanical workshop, storage facilities, chemical lab, electronic workshop, assembly hall, computer and networking, library, canteen, sleeping rooms, conference rooms, headquarters, administration. Special care has been given to the development of structures, instrumentation, procedures and training activities in matter of safety, of the users and of the citizens, and environmental impact. A number of outreach activities and visits to the lab are systematically organised by a dedicated Service. Personnel (physicists, engineers, technicians, administration) include a permanent staff of 64 and presently 23 non-permanent positions.

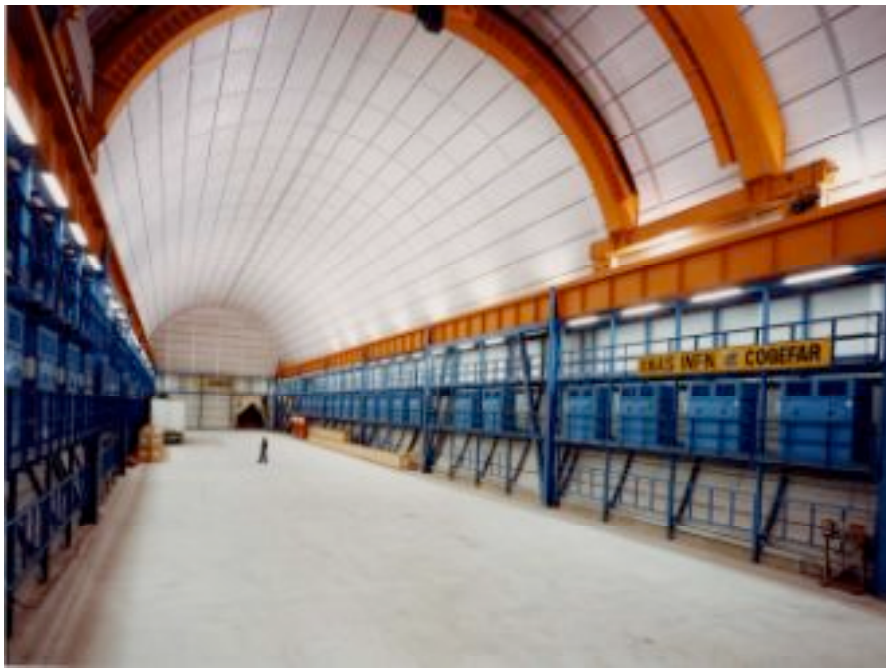
Fig. 2 Hall C of LNGS

**Number of scientific users**: 752 from 26 countries

**Type of access. Interference with nearby activities**

Horizontal through the free way tunnel. Access regulated by a protocol agreement with the free way. Special transports need permission. The Laboratory coexists with the freeway structure and with a National Park.

**Rock coverage (metres of rock), muon flux, neutron flux, radon activity (average with ventilation)**

Rock overburden: 1400 m (3.2 km w.e.)

Neutron flux: total $3.78\times10^{-2}$ $s^{-1}m^{-2}$. <0.05 eV = $1.08\times10^{-2}$, 0.05 eV – 1 keV = $1.84\times10^{-2}$, 1 keV - 2.5 MeV = $0.54\times10^{-2}$, >2.5 MeV = $0.32\times10^{-2}$

Muon flux = $3 \times 10^{-4}$ $m^{-2}$ $s^{-1}$

Radon in the air 50-120 $Bq/m^3$, less at the experiments, when needed

**Ventilation power (time to change one lab volume).** One lab volume in 3.5 hours.

**Installed electrical power and available heat exhaust per unit time**

Power: 3 000 kW; maximum heat exhaust rate: ≈1 900 kW.

**Future. MODULAr** is a proposal for a modular set of liquid Ar TPCs of the ICARUS technology for an active volume of the order of 20 kt to be located in a new facility at shallow depth (1.2 km w.e.) 10 km off-axis from the CNGS beam and of an improved neutrino source at CERN with 1.6 MW beam power.

# LSC. Laboratorio Subterráneo de Canfranc (Spain)

The first underground facility under the Pyrenees, close to a dismissed railway tunnel, was created in the 1980s by A. Morales and the Nuclear and High-Energy Physics Department of the Saragossa University. Taking profit of the excavation of a parallel road tunnel, the new laboratory was later built. The underground structures have been completed in 2005. However, more recently a few design and construction defects have emerged and the necessary reparation works are under way, to be completed by autumn 2008.

LSC is managed by a Consortium between the Spanish Ministry for Education and Science, the Government of Aragon and the University of Saragossa. The surface building has been funded and is presently being designed. It will contain headquarters, administration, library, meeting room, offices, laboratories, storages and a mechanical workshop, safety structures and management, for a total of approximately 1500 m$^2$. A dozen of employees are being hired. In the process of approval of the proposals, an international Scientific Committee advises the Director,

**Director**: Alessandro Bettini; Bettini@pd.infn.it
**Web address**: http://ezpc00.unizar.es/lsc/index2.html

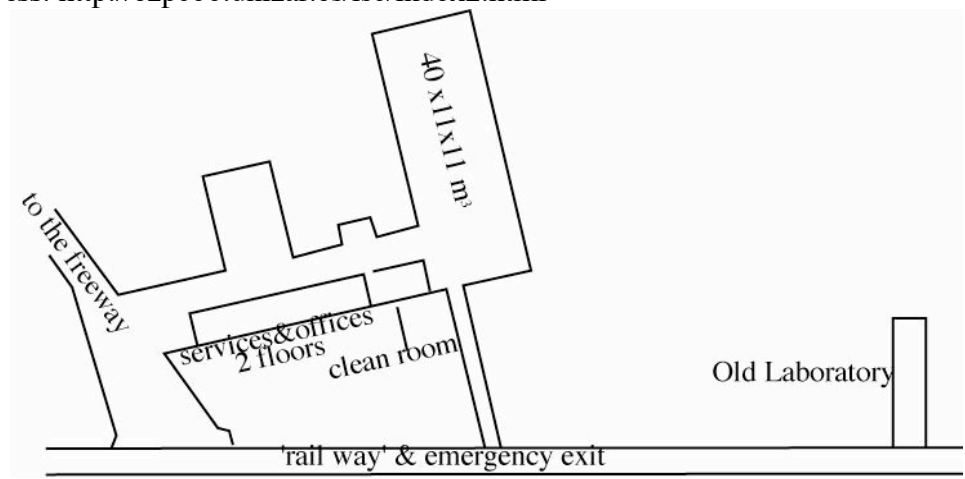

Fig. 1. The underground structures of LSC

**Underground area available for experiments (including the already occupied one)**:
Hall A; 40×15×12 (h) m$^3$; Hall B: 15×10×8(h) m$^3$; Clean room 45 m$^2$; Services 215 m$^2$. Total ≈ 1000 m$^2$. Old lab: 100 m$^2$

**Status of the occupancy of the above area and dates foreseen for the conclusions of the experiments**

The experimental program is being defined. Preliminary Expressions of Interest have already been collected. These include double beta decay search with SuperNEMO, dark matter search with ArDM, with a liquid Ar TPC, SIMPLE with a super-heated fluid and the ultra-cryogenic ULTIMA and a test of gravity. More proposals are welcome. ANAIS and Rosebud, presently active in the old lab will be transferred in the new one.

**Structure of the surface campus, general services provided to the users, number of**

**employees of the lab**

The building, presently being designed, will contain headquarters, administration, library, meeting room, offices, laboratories, storages and a mechanical workshop for a total of approximately 1500 m$^2$. A dozen of employees are being hired.

**Type of access. Interference with nearby activities**

Horizontal, two tunnels (freeway and dismissed railway/safety tunnel). Entrance must be communicated to the freeway tunnel control

**Rock coverage (metres of rock), muon flux, neutron flux, radon activity (average with ventilation)**

Maximum rock coverage: 850 m (2.4 km w.e.); muon flux $2\times10^{-3}$ - $4\times10^{-3}$ m$^{-2}$ s$^{-1}$ depending of the location; neutron flux $2\times10^{-2}$ m$^{-2}$ s$^{-1}$; Radon in the air: 50-80 Bq/m$^3$

**Ventilation power (time to change one lab volume)**

11 000 m$^3$/h; time to change one volume of the laboratory is 40'.

**Installed electrical power and available heat exhaust per unit time**

Electrical power: 500 kW; exhaust capacity > 500 kW.

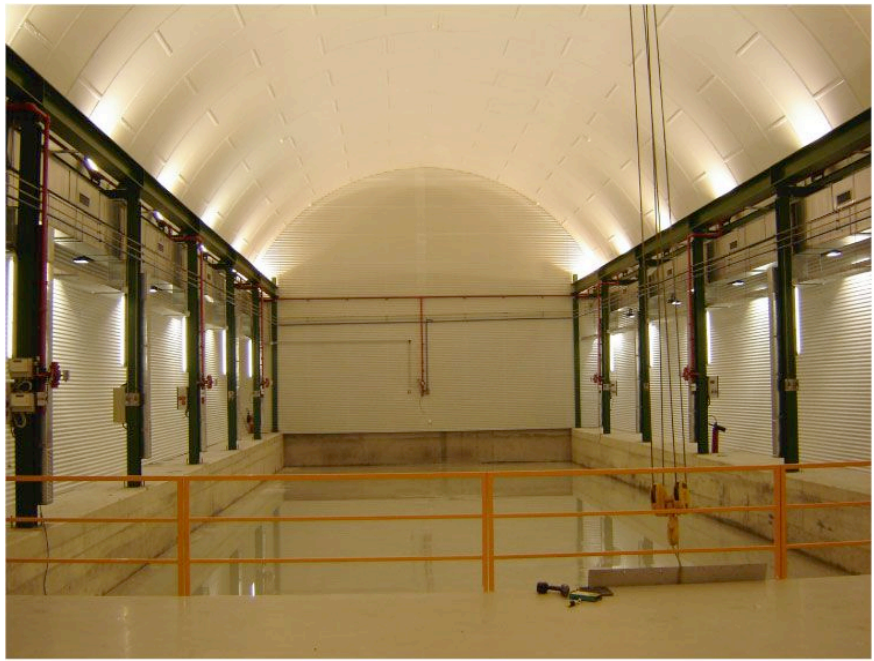

Fig. 2 Hall A of LSC

# LSM. Laboratoire Subterrain de Modane (France)

The Laboratory is operated jointly by the CNRS/IN2P3 and CEA/DSM. The excavation of the Laboratory started in 1979 and was completed by 1982, to host a 900 t Fe tracking calorimeter to search for proton decay. The experiment finished in 1988.

**Director**: F. Piquemal piquemal@cenbg.in2p3.fr

**Web address**: http://www-lsm.in2p3.fr/

**Underground area available for experiments (including the already occupied one)**:
400 m$^2$.

**Status of the occupancy of the above area and dates foreseen for the conclusions of the experiments**

The lab is almost full. It remains ~ 15 m$^2$ in the main hall for temporary experiment like BiPo or Sphere TPC. In the annexes, there is still place for few Ge detectors.

There are 2 larger experiments NEMO 3 (double beta decay) and EDELWEISS (dark matter), they should run at least up to 2010.

**Structure of the surface campus, general services provided to the users, number of employees of the lab**

100 m$^2$ for offices, a warehouse of 150 m$^2$ (with small workshop) and a flat for 3 users.

The lab provides cars to the users to access to the lab.

8 technicians and engineers and 1 post doc in LSM.

**Number of scientific users**: about 100

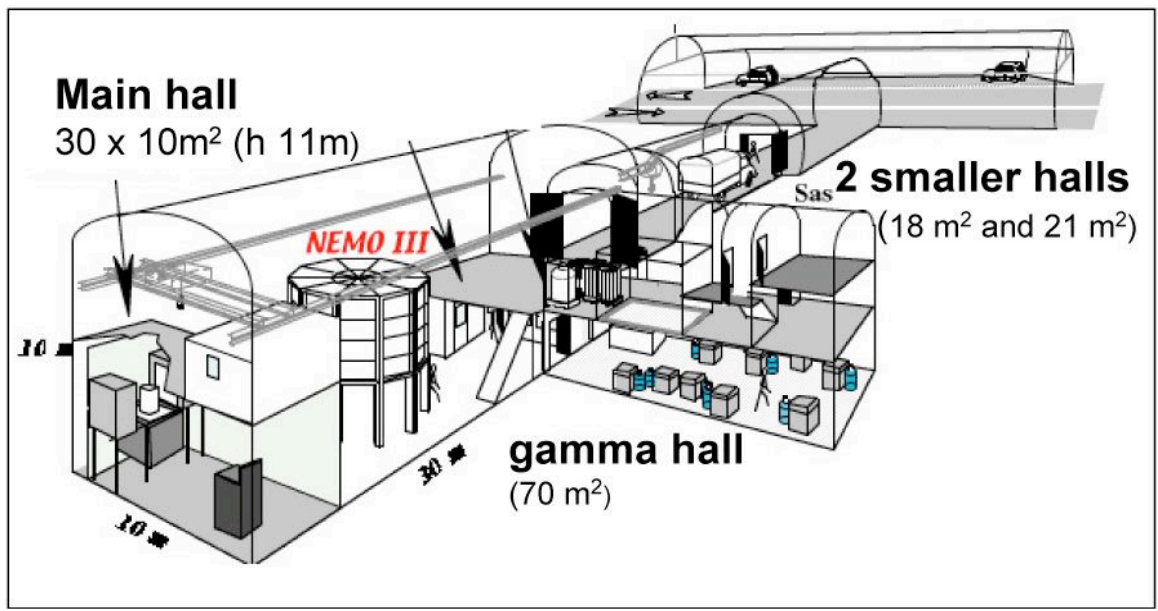

Fig. 1. The structure of LSM, with the present experiments

**Type of access. Interference with nearby activities**

Horizontal access through the Frejus roadway tunnel, at any time. In case a car or a lorry has to go into the laboratory, intervention of the tunnel control is needed to stop the traffic at the level of the lab during the time of the entrance or exit of the vehicle from the lab.

**Rock coverage (metres of rock), muon flux, neutron flux, radon activity (average with ventilation)**

1700 m of rock (4.8 km w.e.), muons: $4.7 \times 10^{-5}$ m$^{-2}$s$^{-1}$; neutrons: $5.6 \times 10^{-2}$ m$^{-2}$s$^{-1}$ (further work in progress); radon: 15 Bq/m$^3$ (antiradon factory, production of 150 m$^3$ of air with 10 mBq/m$^3$)

**Ventilation power (time to change one lab volume)**

1.5 lab volumes/hour

**Installed electrical power** 400 kVA

**Future.**

**The Ulisse project** profits of the opportunity given by the construction of a new tunnel approved by the French and Italian Governments to increase the safety conditions of the tunnel traffic. The Ulisse project foresees two large halls: hall A of 100×24 m$^2$, hall B of 18×50 m$^2$. Hall B will provide an extremely low background environment by surrounding its central volume with a water shield and by artificially producing a very low Rn content atmosphere (0.1 mBq/m$^3$).

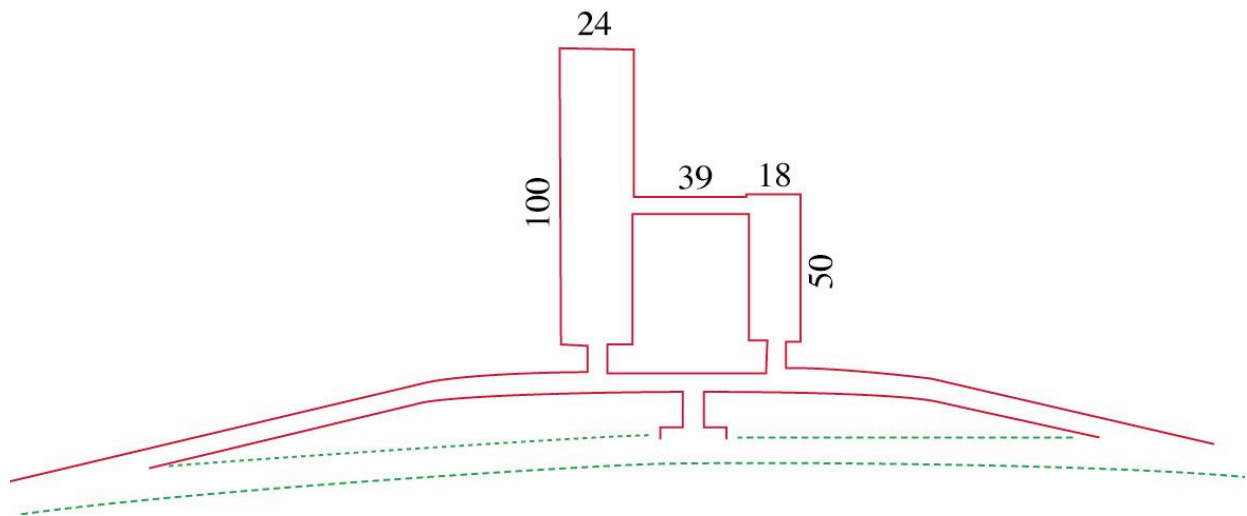

Fig. 2. The two new halls as foreseen in the Ulisse Project and the new Frejus safety tunnel

# SUL. Solotvina Underground Laboratory. INR of UNAS (Ukraine)

The Laboratory was constructed in 1984 under the leadership of Yuri Georgievich Zdesenko by the Lepton Physics Department (LPD) of the Institute for Nuclear Research (Ukrainian National Academy of Sciences) in a salt mine.

**Director**: Fedor Danevich (head of the Lepton Physics Department responsible for the Solotvina Laboratory)

**Web address**: http://lpd.kinr.kiev.ua/LPD_SUL.htm

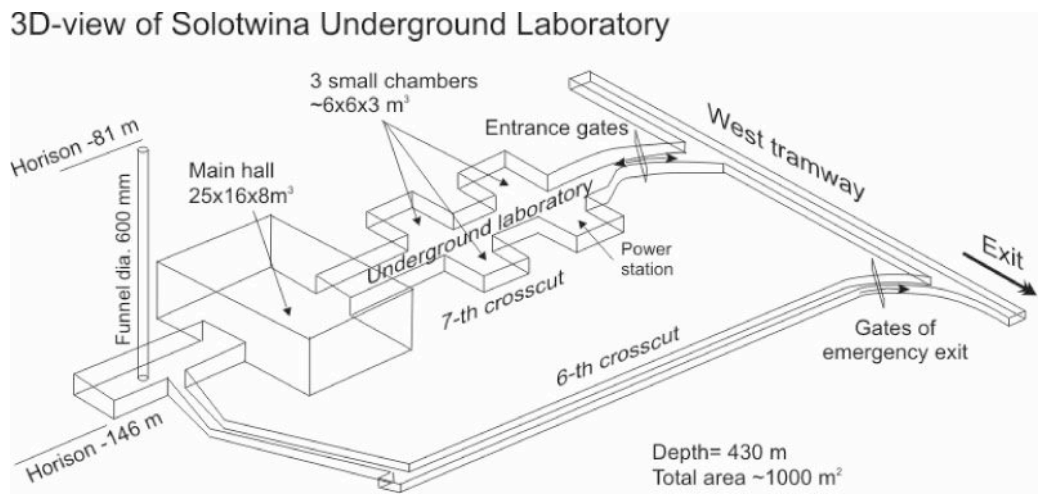

Fig. 1. The underground structures of SUL

**Underground area available for experiments (including the already occupied one)**:
Main hall: 25×18×8(h) m$^3$; 4 chambers 6×6×3(h) m$^3$. Total area: ≈1000 m$^2$

**Status of the occupancy of the above area and dates foreseen for the conclusions of the experiments.** ≈30% of area is occupied; 1.5-2 month

**Structure of the surface campus, general services provided to the users, number of employees of the lab.**
Surface office including three double living rooms; staff: 14 persons: technicians and engineers.

**Number of scientific users**:
11 researchers and PhD students of the Lepton Physics Department of the Institute for Nuclear Research (Kiev).

**Type of access. Interference with nearby activities**
Vertical access by mine cage. Personal of the Underground Lab should take into account time-table of the Solotvina salt mine. Practically no problem to access 24 hours per day. Access is partially limited in Sunday.

**Rock coverage (metres of rock), muon flux, neutron flux, radon activity (average with ventilation)**
430 m in a salt mine (≈1 km w.e.); muon flux: $1.7\times10^{-2}$ m$^{-2}$ s$^{-1}$; neutron flux $<2.7\times10^{-2}$ m$^{-2}$ s$^{-1}$; Radon concentration in air is 33 Bq m$^{-3}$; gamma background is 30-50 times lower than that in a typical surface laboratory; temperature 23±1 C.

**Ventilation power (time to change one lab volume).** Mine ventilation system (7 minutes)

**Installed electrical power.** 65 kW, 4 kW UPS

**Present projects:**

- Advancement of the $^{116}$Cd $\beta\beta$ experiment (new ~1-2 kg $^{116}$CdWO$_4$ higher quality crystal scintillators, improvement of the set-up and data acquisition). The experiment scheduled to start in the beginning of 2008.
- R&D for SuperNEMO $\beta\beta$ project
- R&D of radiopure crystal scintillators for $\beta\beta$ and dark matter experiments: CdWO$_4$, CaWO$_4$, ZnWO$_4$, PbWO$_4$, CaMoO$_4$, new molybdates

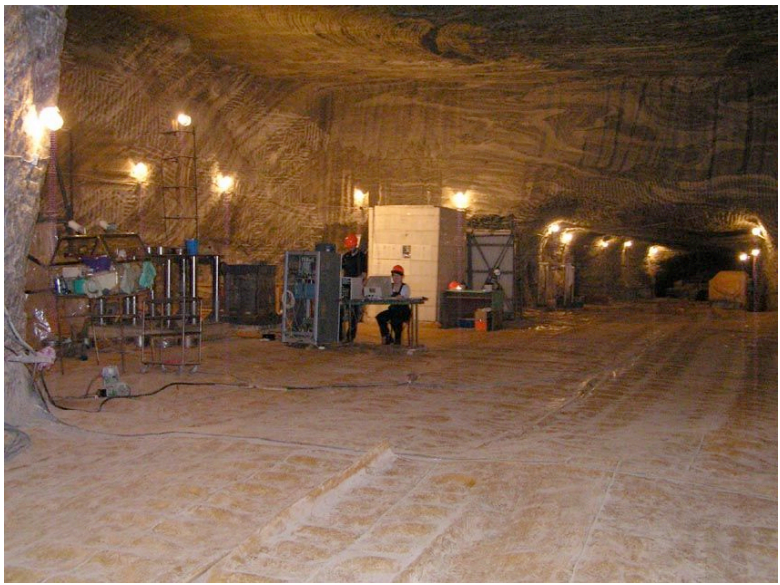

Fig. 2. Solotwina Underground Laboratory

# ASIA

Two laboratories are active in Japan, a larger and a smaller one, one is present in Korea and one is being has been approved in India. All of them are under a mountain with horizontal access. They are listed in alphabetical order

## Kamioka Observatory (Japan)

The Kamioka Observatory is operated by the Institute for Cosmic Ray Research, University of Tokyo. It was established in 1983 by M. Koshiba as Kamioka Underground Observatory. The original purpose of this observatory was the KamiokaNDE experiment, to which Super-Kamionkande followed. The present facilities have been designed for Super-K, the largest existing underground experiment. Recently, a process of enlargement of the underground facilities has started, in order to accommodate more experiments. The proposals are dealt with case by case, as yet. The KamLAND experiment is operated by the Neutrino Centre, Tohoku University, directed by Prof. Kunio Inoue inoue@awa.tohoku.ac.jp.

**Director**: Yoichiro Suzuki; suzuki@icrr.u-tokyo.ac.jp
**Web address**: http://www-sk.icrr.u-tokyo.ac.jp/index_e.html

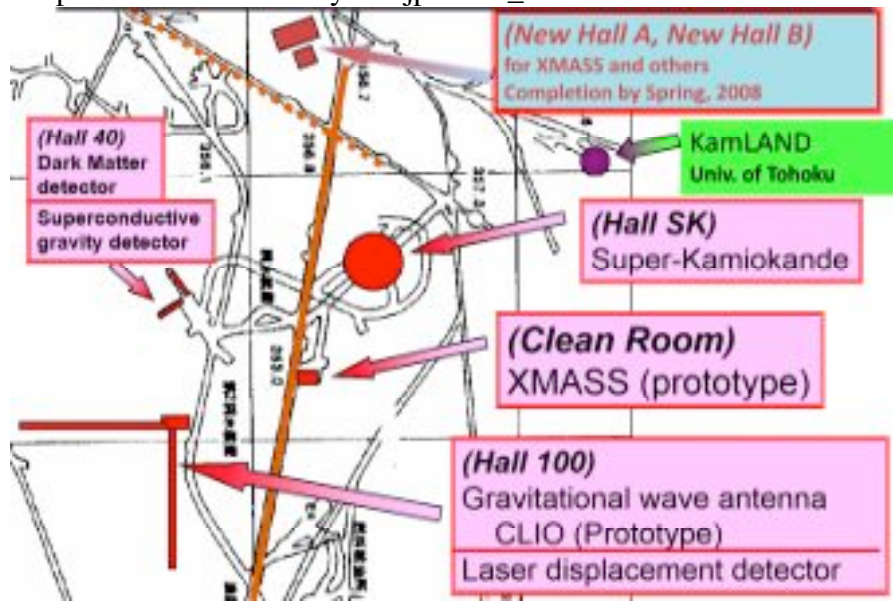

Fig. 1. The underground facilities at Kamioka

**Underground area available for experiments and dates foreseen for the conclusions of the experiments**

Hall SK (50 m diameter), hosting Super-Kamiokande. To be continued another 15 years, at least.
Clean room ($10 \times 5$ m$^2$). XMASS prototype.
Hall 40 (L-shape, 40 m $\times$ 4 m arm). Purification tower for XMASS. NEWAGE, Dark matter, Kyoto Univ. To be terminated by 2008. Superconductive gravity detector
Hall 100 (L-shape, 100 m $\times$ 4 m arm). CLIO: prototype of Gravitational Antenna. To be terminated

in 2013. Laser displacement detector.

Hall A. New (15×21 m$^2$). XMASS 800 kg. To be occupied till 2012. Space for another experiment available.

Hall B, new (6×11 m$^2$). CANDLE; double beta decay, Osaka Univ. To be occupied till March 2012.

Small areas available in the dismissed mine.

**Structure of the surface campus, general services provided to the users, number of employees of the lab**

Buildings for offices and computer facilities. 13 scientists, 2 technical support units, one for administration.

**Number of scientific users**: more than 200

**Type of access. Interference with nearby activities**

Horizontal by car. The mine is no more active

**Rock coverage (metres of rock), muon flux, neutron flux, radon activity (average with ventilation)**

Coverage is 1000 m, corresponding to 2.7 km w.e.. Muon flux = $3 \times 10^{-3} \mu$ m$^{-2}$s$^{-1}$

Neutrons flux; thermal $8.25 \pm 0.58 \times 10^{-2}$ m$^{-2}$s$^{-1}$; non-thermal $11.5 \pm 1.2 \times 10^{-2}$ m$^{-2}$s$^{-1}$

Rn in the air: few Bq/m$^3$ with normal ventilation

**Ventilation power (time to change one lab volume).** 3000 m$^3$/h

**Installed electrical power and available heat exhaust per unit time**

750 kW. Water-cooling system can handle all the heat exhaust

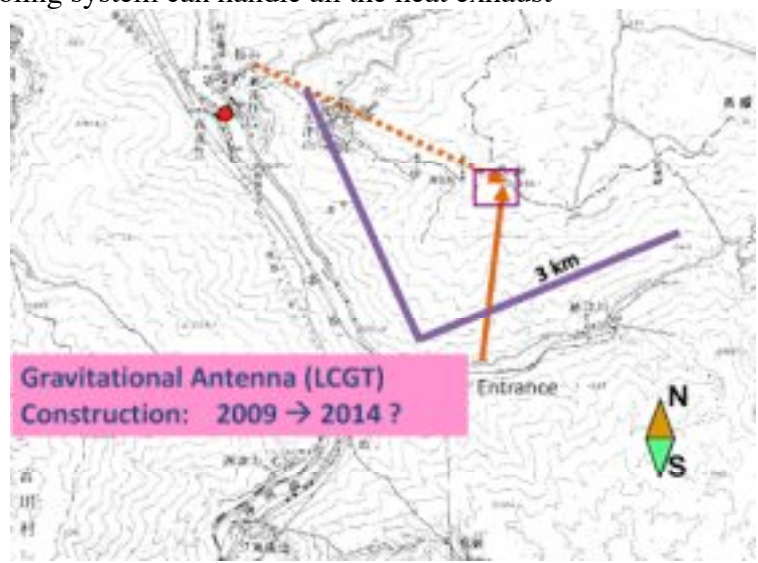

Fig. 2. The site planned for LCGT

**Future.** New Halls A and B under construction. No plan for more cavities, at present. Budget request for the underground gravitational antenna LCGT has been submitted (see Fig. 2).

# OTO-Cosmo Observatory (Japan)

**Director**: Hiro Ejiri; ejiri@pop07.odn.ne.jp

**Web address**: http://wwwkm.phys.sci.osaka-u.ac.jp/info/syoukai/oto-e.html

**Underground area available for experiments (including the already occupied one)**:

≈100 m$^2$. Lab. 2 (50 m$^2$) hosts ELEGANT V on 0ν2β decay of $^{100}$Mo and MOON-1 on dark matter search with NaI. Lab. 1 (33 m$^2$) hosts ELEGANT VI on 0ν2β decay of $^{48}$Ca and dark matter search with CaF$_2$.

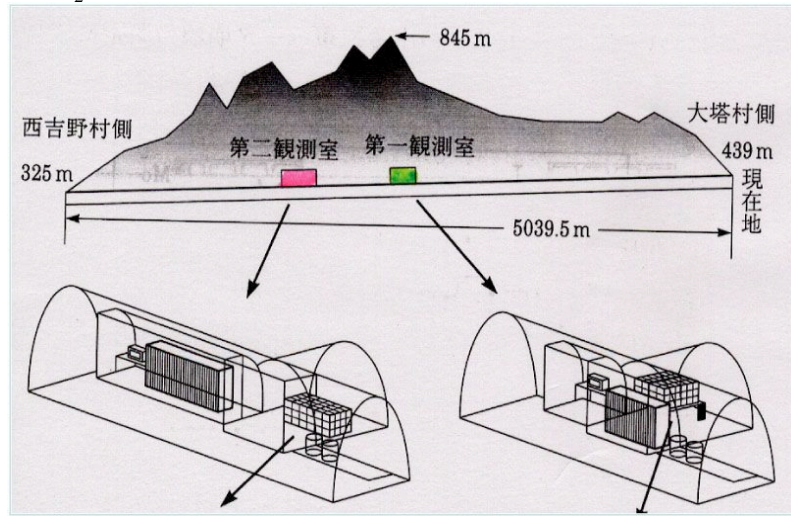

Fig. 1. The OTO-Cosmo underground structures

**Number of scientific users**: ≈20

**Type of access. Interference with nearby activities**

Horizontal through non-used railway tunnel

**Rock coverage (metres of rock), muon flux, neutron flux, radon activity)**

Rock coverage: 470 m (1.4 km w.e.); muon flux: $4\times10^{-3}$ m$^{-2}$s$^{-1}$; neutron flux: $4\times10^{-2}$ m$^{-2}$s$^{-1}$; radon concentration: 10 Bq m$^{-3}$ (in "radon free" containers)

**Ventilation power:** Natural wind through the tunnel

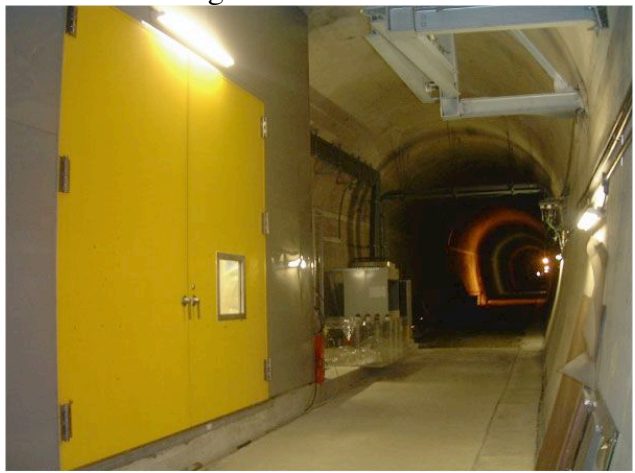

Fig. 2. The OTO-Cosmo Laboratory

# Y2L

Operated by the Dark Matter Research Centre (DMRC) of Seoul National University under the agreement between SNU and the Korea Midland Power Co. Ltd. who owned the YangYang Pumped Storage Power Plant

**Director**: Sun Kee Kim skkim@hep1.snu.ac.kr

**Web address**: dmrc.snu.ac.kr

**Underground area available for experiments (including the already occupied one)**:

Current space: 100 m$^2$

Planned space: 800 m$^2$ (funding not available yet)

**Status of the occupancy of the above area and dates foreseen for the conclusions of the experiments** (including the already occupied one)

- Mostly occupied by the Korea Invisible Mass Search (KIMS) experiment - currently data taking for WIMP search with CsI(Tl) crystal detectors. KIMS installed 100 kg crystals this year.
- R&D run for double beta decay experiment
- HPGe detector for background measurement

**Structure of the surface campus, general services provided to the users, number of employees of the lab**

- 100 m$^2$ space for office space, computing and detector test facility
- one room with kitchen and bath for lodging of ~ 2 people

**Number of scientific users**: ~ 30

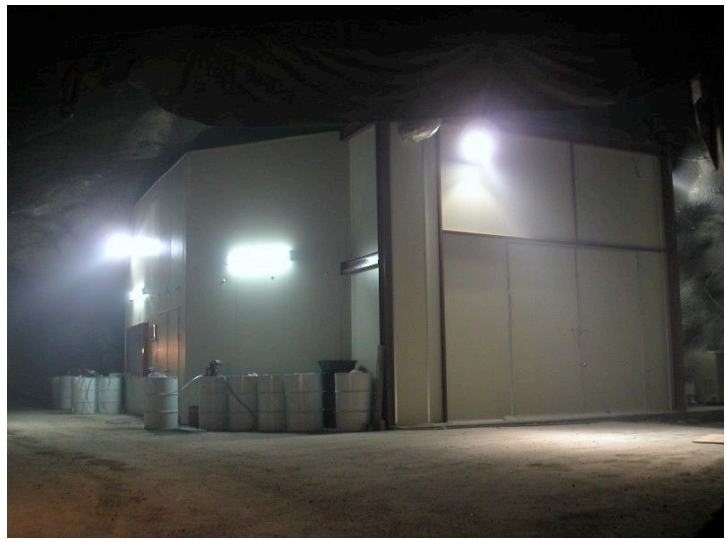

Fig. 1 The Y2L Laboratory

**Type of access. Interference with nearby activities**

- horizontal access by car (2 km paved road from the entrance of the tunnel)
- access at any time (for registered users)
- The lab utilizes the space in the tunnel for Yangyang Pumped Storage Power Plant, currently in

operation. Researchers are requested to obey the safety and security regulation of the Plant.

**Rock coverage (metres of rock), muon flux, neutron flux, radon activity (average with ventilation)**

- Rock overburden : 700 m, ~ 2 km w.e.
- muon flux: $2.7 \times 10^{-3}$ m$^{-2}$s$^{-1}$
- Neutron flux: $8 \times 10^{-3}$ m$^{-2}$s$^{-1}$ for $1.5 \text{MeV} < E_n < 6.0$ MeV
- Radon activity: 40-80 Bq/m$^3$

**Ventilation power (time to change one lab volume)**

- ventilation of the whole tunnel is controlled by the power plant

**Installed electrical power and available heat exhaust per unit time**

- in the lab. ~ 40 kVA for current usage
- can be expandable to as desired, in principle

# INO

The India based Neutrino Observatory is the project, approved by the Government, to create an underground laboratory in southern India. To be organized with international laboratory standards with a Scientific Advisory Committee, services to the users, environmental, safety, security and outreach activities.

**Director**: Naba Mondal; nabak.mondal@gmail.com

**Web address**: http://www.imsc.res.in/~ino/

**Underground area available for experiments (including the already occupied one)**:
Lab1: 26×135×25 (h) m$^3$, Lab 2: 53.4×12.5×8.6 (h) m$^3$ + connecting tunnels and services

**Status of the experiments**

Main foreseen experiment: ICAL: 50 kt magnetized Fe tracking calorimeter for atmospheric and very long base-line accelerator neutrinos. It will occupy only a fraction of Lab1

**Structure of the surface campus, general services provided to the users, number of employees of the lab**

1400 m$^2$ building for administration, offices, shops, etc.; 2750 m$^2$ building with lecture hall and guest house; residential complex with 20 quarters. Personnel: 50-100

**Type of access. Interference with nearby activities**

Horizontal through dedicated 2 km tunnel

**Rock coverage.** 1400 m rock overburden (3.5 km w.e.)

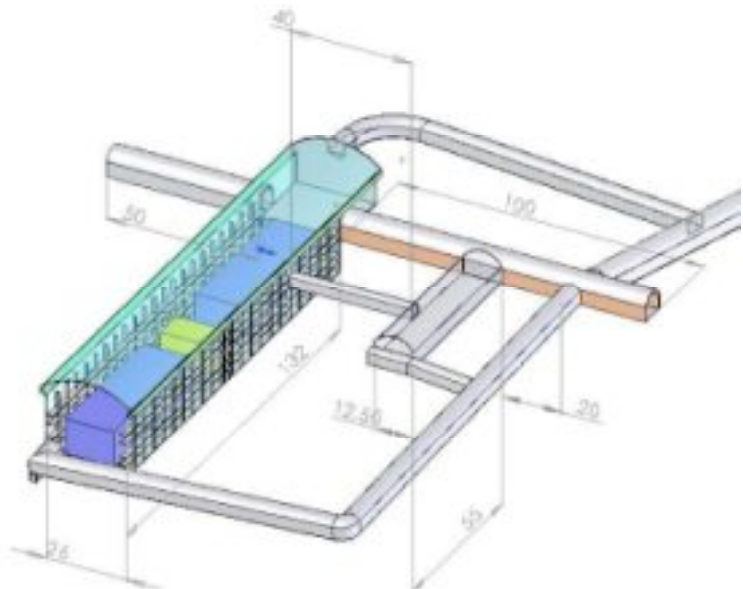

Fig. 1. A view of the future underground structures of INO

# NORTH AMERICA

In Canada the site of the SNO experiment is being developed into a larger underground laboratory. In the US there is one laboratory, SOUDAN. In 2004 NSF launched a solicitation for a full-fledged underground scientific laboratory 'Deep Underground Science and Engineering Laboratory = DUSEL'. In spring 2007 NSF selected the proposal to fund the development of a complete project at the Homestake mine site.

All laboratories are in a mine or a former mine site.

## SNO-laboratory. Sudbury (Northern Ontario, Canada)

**Director**: Tony Noble; potato@snolab.ca

**Web address**: http://www.snolab.ca/   http://www.sno.phy.queensu.ca/

The SNO experiment has completed its life and its cavity is now being freed for further experimental activity. Moreover, new structures are being excavated.

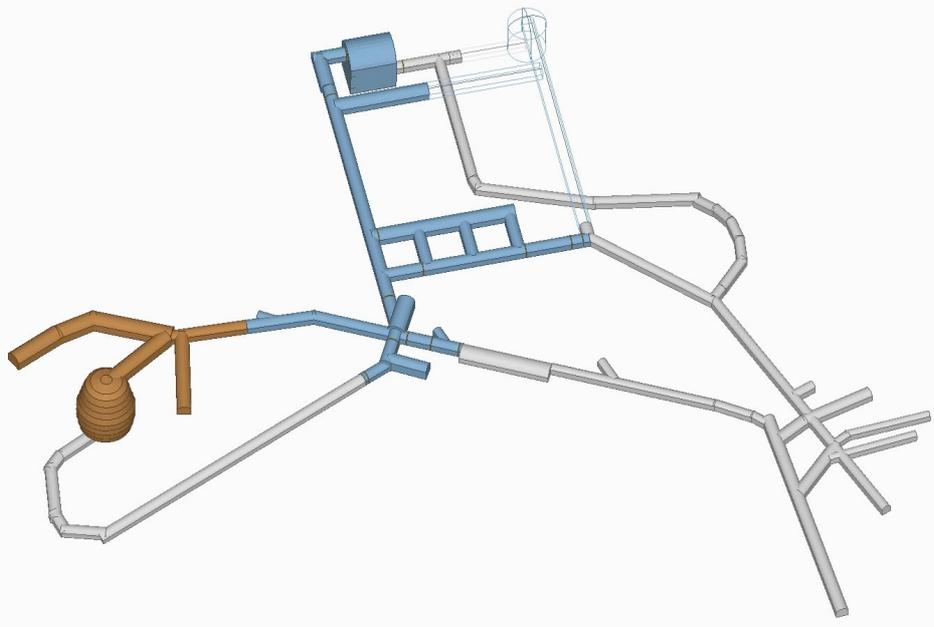

Fig. 1. The former SNO cavity is on the left, the new Laboratory in the centre. Dimmed the further project

**Underground area available for experiments (including the already occupied one)**:

New structure are under construction: a Main hall of volume $18 \times 15 \times (15$ to $19.5$ height$)$ m$^3$, a service hall of about 180 m$^2$ and a number of narrow (6-7 m) volumes called "ladder labs". The construction of a further structure, called cryopit, has been recently approved. This hall is designed to cope with the safety issues surrounding large volumes of cryogenic fluids. The total area will be 7 215 m$^2$, of which 3 055 m$^2$ available for the experiments, the total volume will be 46 648 m$^3$ of which 29 555 m$^3$ available for the experiments. The access will be vertical, through the shaft of the working mine, available daily. All the laboratory will be clean, class 1500.

**Status of the occupancy of the above area and dates foreseen for the conclusions of the**

**experiments**

PICASSO, searching for dark matter (2 kg) with the super-heated bubbles technique, is already running. SNO+ will be hosted in the former SNO cavity; it will be based on liquid scintillator for low energy solar neutrinos, geoneutrinos and double beta decay by dissolving $^{150}$Nd in the liquid. Dark Matter search includes DEAP/CLEAN with noble liquids, which is getting ready to install prototype. A LoI from superCDMS with bolometers has been received. More LoIs are expected to be reviewed by the Experimental Advisory Committee.

**Structure of the surface campus, general services provided to the users, number of employees of the lab**

Surface facilities include: a 3 159 m$^2$ building which includes: class 1000 clean room laboratories, staging and assembly areas for experiments, office space (60 users), meeting rooms, control rooms, IT server room, emergency generator, high speed network link off site, high speed network link surface/underground. The structures can handle more than 100 users at a time.

30 full-time staff.

**Type of access. Interference with nearby activities**

The access will be vertical, through the shaft of a working mine, available daily. All the laboratory will be clean, class 1500.

**Rock coverage** is 2000 m under flat surface (6.01 km w.e.) Muon flux $3 \times 10^{-6}$ m$^{-2}$s$^{-1}$, thermal neutron flux = $4.7 \times 10^{-2}$ m$^{-2}$s$^{-1}$, fast neutron flux = $4.6 \times 10^{-2}$ m$^{-2}$s$^{-1}$, Radon in the air: 120 Bq/m$^3$

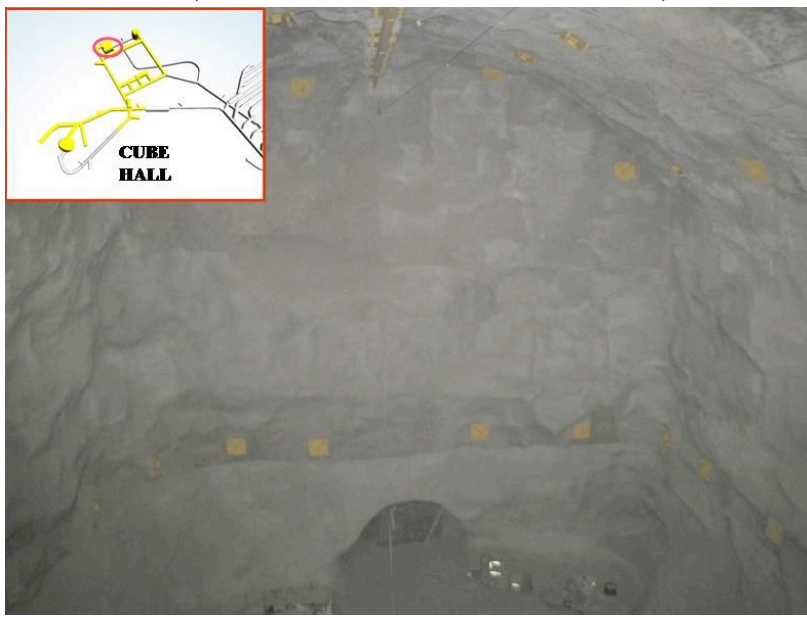

Fig. 2. One of the new 'ladder' laboratories

**Ventilation power (time to change one lab volume)**

Smaller lab spaces: 10 air changes per hour through HEPA filters. Larger lab spaces: 5 air changers per hour through HEPA filters.

**Installed electrical power and available heat exhaust per unit time**

2MW electrical power available; 1MW heat removal available

# SUL. Soudan Underground Laboratory (Minnesota USA)

Director: Earl Peterson eap@physics.umn.edu

**Web address**: http://www.soudan.umn.edu/

**Underground area available for experiments (including the already occupied one)**:

Two halls with associated (short) connecting tunnels:

a. The "Soudan" lab is 72×14×14(h) m$^3$ = 1000 m$^2$

b. The "MINOS" lab is 82×16×14(h) m$^3$ =1300 m$^2$

Total experimental area = 2300 m$^2$

**Status of the occupancy of the above area and dates foreseen for the conclusions of the experiments**

Soudan lab: CDMS II (20×7×10(h) m$^3$) expects to run until 2009 and perhaps utilize the space for testing detectors for a few years more. The emerging low-background counting facility currently occupies 5×5×3 m$^3$, but – if fully funded – will expand to 25×14×14(h) m$^3$. It will run until funding limitations intrude (unknown time frame but extending until at least 2009).

MINOS lab: MINOS occupies 35×16×14(h) m$^3$. It expects to run until 2009 or longer with a 2-year decommissioning period at the conclusion. The high-purity copper fabrication facility occupies 4×6×3(h) m$^3$ and expects to run for at least another two years.

**Structure of the surface campus, general services provided to the users, number of employees of the lab**

The major facility on the surface is a receiving building with a 7.5 t crane and two office areas, kitchen and sanitary facilities. It occupies approximately 650 m$^2$.

The laboratory has a staff of 9, including secretarial and accounting assistance and network and computer maintenance personnel. It is staffed 10 hours/day, 5 days per week, but the staff is on-call during the balance of the time and responds to requests for emergency access.

**Number of scientific users**: MINOS is a collaboration of approximately 200, CDMS II about 60 and the smaller efforts (low-background/copper fabrication) a few (about 5). **Total**: **265**

**Type of access. Interference with nearby activities**

Vertical access via a two-compartment slightly angled shaft. Diameters in excess of 1m and lengths in excess of 10m pose a problem. Access outside normal operating hours is possible. There is an access charge paid to our host institution, Soudan Underground Mine State Park.

Normal laboratory safety requirements are in place. The laboratory coexists in the Soudan mine with an historic State Park, which offer mine tours during the summer months to the public, and winter tours to school groups. Some tours utilize a visitor's gallery available in the MINOS laboratory, but do not interfere with either operation or installation of equipment. Access for prolonged equipment haulage beyond two hours in the morning and after tours end in the afternoon is difficult in the summer months. There is no active mining activity.

**Rock coverage, muon flux, neutron flux, radon activity (average with ventilation)**

The overburden is 700m of Soudan rock (2 km w.e.). The muon flux is $2\times10^{-3}$ m$^{-2}$ s$^{-1}$. The neutron interaction rates are approximately 10 kg$^{-1}$ d$^{-1}$ (from U/TH, low energy) or 0.01 kg$^{-1}$ d$^{-1}$ (muon generated in the rock). The radon concentration is seasonal, varying from 300 Bq/m$^3$ in the winter to 700 Bq/m$^3$ in the summer.

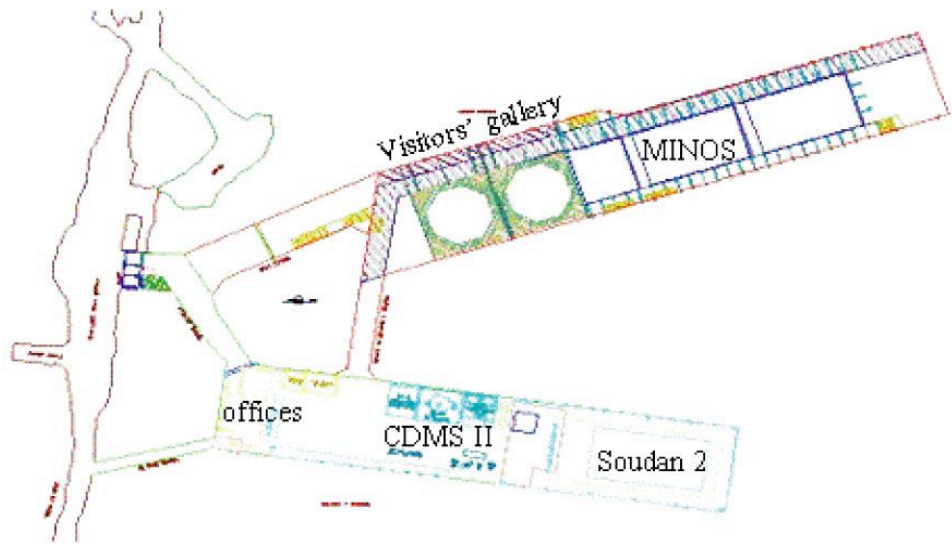

Fig. 1. Map of the underground halls and corridors

**Ventilation power (time to change one lab volume)**
The Soudan mine has natural ventilation – about 550 m$^3$/h for the level of the laboratories. Half of this is diverted to ventilate the MINOS and Soudan spaces. This results in a complete air change every 110'.

**Installed electrical power and available heat exhaust per unit time**
Currently one looped 600 kVA line serves both the Park and the laboratories underground. Approximately 370 kVA is utilized (at peak) by the Park, mostly for pumping. Of the 230 kVA available, about 180 kVA is currently used by MINOS, CDMS II and general laboratory operation. There is both a plan to add another 600 kVA loop for laboratory expansion capability, and the financing to install it. It can be implemented at need.

The cooling capacity is currently 400 kW. It is approximately somewhat less than 50% utilized

# DUSEL. (South Dakota USA)

After a long and complex process, in spring 2007, NSF has selected amongst several proposals the Homestake mine as the site in which the Deep Underground Science (physics, biology, and geology) and Engineering Laboratory (DUSEL) should be designed.

Expected NSF funding for the facility ≈250 M$ + contribution to initial set of experiments ≈250 M$. In addition, to prepare the site, South Dakota provides 46 M$ on its own and $70M from a donation from T. D. Sanford, a SD banker + 50GB/s connection (8 M$).

NSF is funding a 3-year effort (15 M$ expected) to design the facility in preparation for the FY011 funding.

For now, the facility is a SD State Laboratory, funded primarily from SD controlled money.

Water inflow in the mine = 1.2 Mt/yr. Pumps stopped in spring 2003. Rehabilitation work started in Jan. 2007. 4850 ft level secured, with pumps at 5300 ft. 7400L water @ about 7 Mpa.

Operating costs being discussed; first estimate ≈25 M$ mainly for direct support to experiments.

**Director**: Kevin Lesko KTLesko@lbl.gov

**Web address**:

**Underground area available for experiments (including the already occupied one)**:

Laboratory spaces will be built separately for biology, geology and physics. Service and R&D structures (e.g. electroforming) will be available for physics at 100 m deep level. Two main campuses are foreseen for physics about 1450 m (4850' L) and 2200 m (7400' L) deep. Each will contain a number (4 and 5 respectively) of standard modules plus service areas (Fig. 2). An experimental module is 50×20×15 (h) m$^3$ on two floors (Fig. 1).

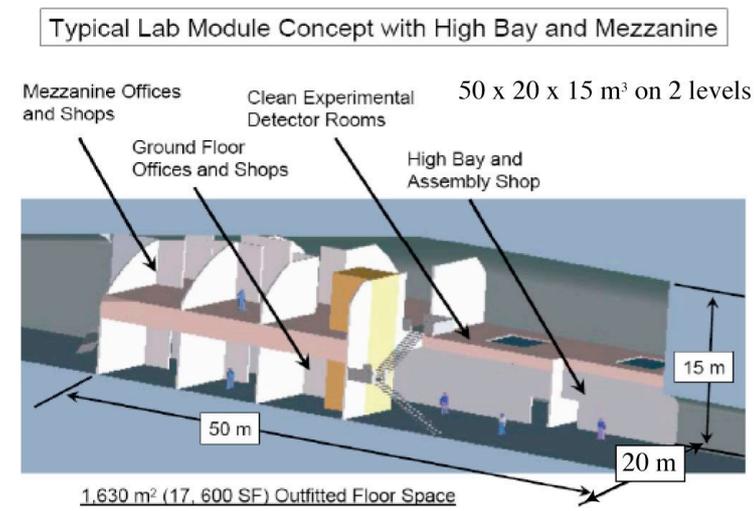

Fig. 1

Present schedule foresees the excavation of the first two 4850L modules in 2008-2009 and of the second two in 2009-2010. The construction of 7400L requires NSF and Congress approval. Its development is estimated in 2012 – 2014.

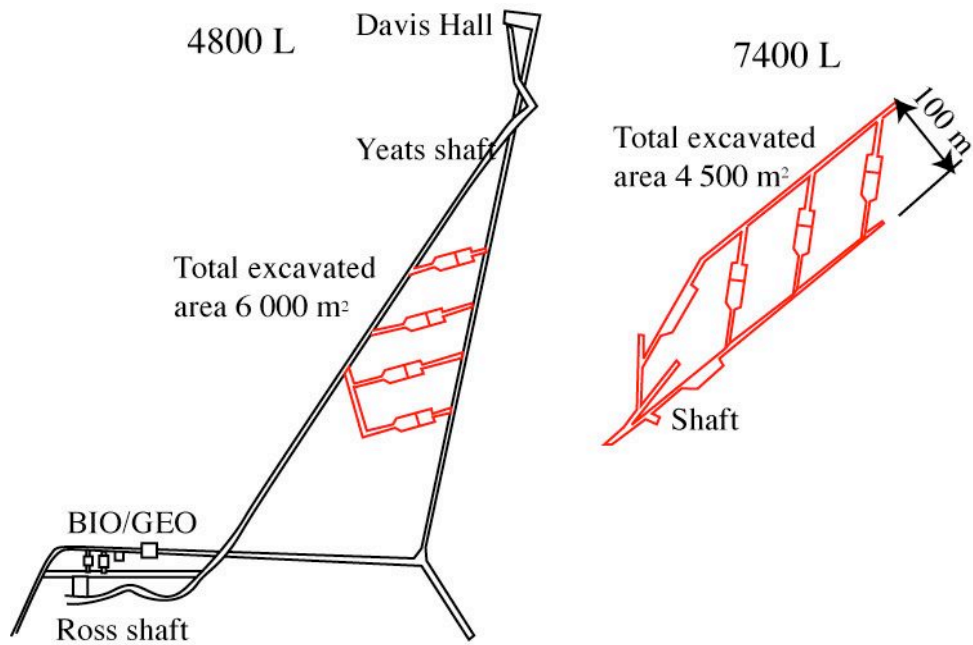

Fig. 2

**Structure of the surface campus, general services provided to the users, number of employees of the lab**

Surface campus ≈ 10 000 m$^2$, will be obtained by rehabilitated existing buildings, close to the access lifts. Will host offices, support, laboratories, etc.

A major science education centre (funded by Sanford) is foreseen.

Initial staff is estimated to be ≈30-50, to increase to 100-150, as State lab will evolve to DUSEL

Will include both professional safety and environmental staff

**Scientific users**

The community will include physicists, biologists, geologists and engineers.

NSF is expected to announce a call for experimental proposals soon

Scientific Committee established on National basis in 2006

**Type of access.**

Access will be vertical through dedicated shafts. Mine is no longer operational